\documentclass[lettersize,journal]{IEEEtran}
\usepackage{multirow}
\usepackage{amsmath,amsfonts}
\usepackage{algorithmic}
\usepackage{algorithm}
\usepackage{array}
\usepackage[caption=false,font=normalsize,labelfont=sf,textfont=sf]{subfig}
\usepackage{textcomp}
\usepackage{stfloats}
\usepackage{verbatim}
\usepackage{graphicx}
\usepackage{subcaption}
\usepackage{booktabs}
\usepackage{caption}
\usepackage{cite}
\usepackage{tabularx}
\begin{document}

\title{Causal-Inspired Multi-Agent Decision-Making via Graph Reinforcement Learning}

\author{
    Jing Wang, 
    Yan Jin,~\IEEEmembership{Member,~IEEE,}
    Fei Ding,
    Chongfeng Wei*,~\IEEEmembership{Senior Member,~IEEE,}

\thanks{*Chongfeng Wei is the corresponding author.}
\thanks{Jing Wang, Yan Jin and Chongfeng Wei are with the School of Mechanical and Aerospace Engineering, Queen’s University Belfast, Belfast, United Kingdom (email: jwang61@qub.ac.uk, y.jin@qub.ac.uk, c.wei@qub.ac.uk) }
\thanks{Fei Ding is with the State Key Laboratory of Advanced Design and Manufacturing Technology for Vehicle, College of Mechanical and Vehicle Engineering, Hunan University, Changsha 410082, China  (email: dingfei@hnu.edu.cn)}
\thanks{Chongfeng Wei is also with James Watt School of Engineering, University of Glasgow, Glasgow, G12 8QQ, United Kingdom (email: chongfeng.wei@glasgow.ac.uk)}}


\IEEEpubid{}

\maketitle

\begin{abstract}
Since the advent of autonomous driving technology, it has experienced remarkable progress over the last decade. However, most existing research still struggles to address the challenges posed by environments where multiple vehicles have to interact seamlessly. This study aims to integrate causal learning with reinforcement learning-based methods by leveraging causal disentanglement representation learning (CDRL) to identify and extract causal features that influence optimal decision-making in autonomous vehicles. These features are then incorporated into graph neural network-based reinforcement learning algorithms to enhance decision-making in complex traffic scenarios. By using causal features as inputs, the proposed approach enables the optimization of vehicle behavior at an unsignalized intersection. Experimental results demonstrate that our proposed method achieves the highest average reward during training and our approach significantly outperforms other learning-based methods in several key metrics such as collision rate and average cumulative reward during testing. This study provides a promising direction for advancing multi-agent autonomous driving systems and make autonomous vehicles' navigation safer and more efficient in complex traffic environments.
\end{abstract}

\section{Introduction}
\IEEEPARstart{W}{ith} the advanced development in autonomous driving technologies, modern transportation has been gradually reshaped, paving the way for safer, more efficient, and environmentally sustainable mobility solutions. Effective decision-making is essential for autonomous vehicles to navigate complex environments, interact with human-driven vehicles (HVs), and respond appropriately to unknown situations. Currently, Graph-based Reinforcement Learning (GRL), incorporating Graph Neural Network (GNN) techniques with Reinforcement Learning (RL), has become a widely adopted approach for decision-making in scenarios involving complex and interdependent interactions among multiple traffic participants. \\
However, several key challenges have to be addressed. Firstly, GRL algorithms often fail to extract sufficient and valuable information from neighboring agents, resulting in data inefficiency and limited scalability \cite{Adibi2023GraphNN}. This limitation arises these methods heavily rely on the characteristics of the training data, which often fail to encompass the full variability encountered in diverse operational scenarios.\cite{10161704}. 
Secondly, GRL algorithms typically construct relationships between vehicles based on observed correlations \cite{Velickovic2017GraphAN,Kipf2016SemiSupervisedCW,Job2023ExploringCL, Jiang2023WhenGN}. Since GNN excels at aggregating and propagating information across graph-structured data by leveraging node features and graph connectivity, they primarily rely on correlation-based relationships \cite{Jiang2023WhenGN}.\\
Therefore, incorporating CDRL into GRL offers a promising solution to address these above-mentioned challenges. By leveraging CDRL, the model is able to disentangle and identify underlying causal factors that govern interactions among vehicles, rather than relying solely on observed correlations. This supports the extraction of features that more accurately model the causal influence of surrounding agents on autonomous decision-making, thereby mitigating data inefficiency, promoting scalability, and enabling autonomous vehicles to learn effectively from fewer interactions \cite{Zeng2023ASO,Deng2023CausalRL}.
Currently, there is no method that effectively combining RL-based algorithms with causal learning algorithms to improve the efficiency and quality of policy generation in multi-agent environments \cite{Grimbly2021CausalMR}. Even though GRL excels at modeling interactions among agents and causal models uncover underlying causal structures, the two models have not unified into a single framework to guide autonomous vehicles in generating more optimal policies in complex multi-agent traffic scenarios.\\
We model the decision-making problem in a multi-agent traffic environment and propose a causality-inspired graph reinforcement learning (CGRL) framework. Our method seeks to learn causally disentangled representations within the Variational Graph Auto-Encoders (VGAE) framework \cite{Kipf2016VariationalGA}. Specifically, we use CDRL to identify and separate causal features from observed data, which are then fed into the designed GRL algorithm to enhance decision-making performance.
To achieve this, we use a GNN-based encoder in VGAE to learn latent representations. Additionally, using information-theoretic methods to extract causal features in the latent space that influence autonomous vehicle decision-making. This process enables the separation of invariant (causal) features from variant (non-causal) ones, based on the premise that invariant representations correspond to causal factors directly related to correct decision-making.\\
Precisely, the contributions of this paper are as follows:\\
1. We propose a CDRL method that can be effectively applied into the VGAE framework.\\
2. This research introduces an innovative multi-agent decision-making framework CGRL that incorporates CDRL into GRL. This framework discerns the causal features which can causally influence the optimal decision-making of autonomous vehicles.\\
3. The innovative CGRL algorithm is implemented and validated in a simulator. Our approach achieves superior performance in the unsignalized intersection scenario, as demonstrated by the results.\\
The following section are organized as follows: 
Section II provides a review of related work on Graph Reinforcement Learning, Causal Reinforcement Learning algorithms, and Causal Disentanglement Representation Learning.
Section III defines an unsignalized intersection scenario and assigns brief identifiers to two kinds of vehicle.
Section IV introduces our developed CGRL algorithm employed in this study.
Section V demonstrates the establishment of experimental scenarios and introduces implementation details.
Section VI analyzes experimental results and provides further discussion.
Section VII presents the conclusion.

\section{Related Work}
\subsection{Graph Reinforcement Learning}
Deep reinforcement learning (DRL) has shown great promise in the domain of autonomous driving. However, as the number of agents increases, their interactions may grow exponentially in complexity. Consequently, DRL-based approaches in multi-agent systems still face several fundamental challenges.
Recently, the integration of GNN with DRL, particularly in multi-agent systems, has gained significant attention in graph-structured environments. GNN are inherently designed to capture topological relationships, making them well-suited for learning the interactions between vehicles in a graph. Additionally, GNN excels at capturing multi-agent relationships compared to other approaches \cite{sai2024}.  Therefore, the combination of DRL and GNN allows for optimizing complex problems while generalizing effectively to unseen topologies \cite{Adibi2023GraphNN}. Recently, a graph convolution-based DRL algorithm was presented that combines Graph Convolution Network (GCN) with Deep Q-Networks (DQN) to achieve improved decision-making in graph-structured environments, such as highway ramping, lane-changing scenarios \cite{Liu2022GraphCD, Yang2022GeneralizedSG, Gao2022MultiAgentDM, Chen2021GraphNN} and intersection \cite{Klimke2022CooperativeBP}. The rise of Graph Attention Network (GAT) is primarily attributed to their attention mechanism, which dynamically assigns importance weights to neighboring nodes rather than relying on fixed aggregation rules. Therefore, some studies proposed GAT-based DRL algorithms \cite{Shi2020EfficientCA,Cai2021DQGATTS, DRL-GAT-SA}.
Despite its greater potential in multi-agent environments, the combination of GNN and DRL still struggles with sample inefficiency \cite{sai2024}.

\subsection{Causal Reinforcement Learning}
Over the past few decades, both causality and reinforcement learning have made significant theoretical and technical advancements independently. However, these two fields have yet to be fully reconciled and integrated. Combining causality with reinforcement learning has the potential to enhance generalization capabilities and improve the sample efficiency of reinforcement learning models. \cite{Zeng2023ASO,Deng2023CausalRL}. Currently, causal reinforcement learning can be broadly classified into two categories. The first category relies on prior causal information, where methods typically assume that the causal structure of the environment or task is provided in advance by experts. The second category, on the other hand, deals with unknown causal information, where the relevant causal relationships must be learned through interaction with the environment to inform the policy. In the context of causal reinforcement learning with given prior causal information, 
Méndez-Molina et al., \cite{MndezMolina2020CausalBQ}  integrated causal knowledge into Q-learning to enhance the agent's ability to learn effectively from its environment. By leveraging this prior knowledge, agents can make informed decisions in complex tasks. 
In \cite{Lu2022EfficientRL}, the study emphasizes the advantages of incorporating prior causal knowledge to guide the learning process in reinforcement learning settings, showing that this approach can improve performance in a comparison of traditional methods that do not account for causal relationships.
However, the complex traffic environment is highly dynamic and involves unpredictable factors, such as road conditions and interactions with other vehicles, making it difficult to fully capture these variables with predefined causal structures. A causal reinforcement learning approach based on unknown causal information can be applied to such traffic scenarios. The authors proposed a runtime method called Counterfactual Behavior Policy Evaluation which focuses on the application of counterfactual reasoning within the context of autonomous driving to enhance decision-making processes \cite{Hart2020CounterfactualPE}. This study introduces a novel framework called Causality-driven Hierarchical RL aimed at improving the discovery of hierarchical structures in RL tasks, particularly in complex environments \cite{Peng2022CausalitydrivenHS}. 

\subsection{Causal Disentanglement Representation Learning}
Currently, capturing the causal structure of relevant variables and extracting causal information to train RL models is a challenging task. 
A traditional method for discovering causal relationships relies on interventions or randomized experiments, which are often too expensive, time-consuming, or even impractical to carry out in many cases. As a result, causal discovery through the analysis of purely observational data has garnered significant attention \cite{Spirtes2000ConstructingBN}. However, current causal discovery methods typically assume that units in the system are random variables and these variables are connected through a causal graph. In real-world complex traffic scenarios, observational data are not always naturally divided into such independent causal units, and the causal discovery approaches are incapable of dealing with high dimensional and complex data \cite{alali:hal-04666466}.\\
To address the issues mentioned in causal discovery, CDRL is proposed. This approach focuses on identifying and isolating independent causal factors within the data\cite{Reddy2021OnCD}. 
In the context of supervised CDRL, a framework called ICM-VAE was proposed to learn causally disentangled representations by utilizing causally related observed labels. This enables the model to leverage supervision from known causal relationships within the data. The authors \cite{Shen2020WeaklySD} introduced a novel method called Disentangled Generative Causal Representation aimed at learning disentangled representations that account for causal relationships among latent variables. In addition, the approach employs weakly supervised information, meaning that it utilizes some level of supervision regarding ground-truth factors and their causal structure without requiring fully labeled datasets. This allows the model to learn effectively even with limited supervision.
For unsupervised manner, a VAE-based CDRL framework was proposed by \cite{Yang2020CausalVAEDR} introducing a novel Structural Causal Model (SCM) layer, enabling the recovery of latent factors with semantic meaning and structural relationships through a causal directed acyclic graph (DAG). The authors proposed a Concept-free Causal VGAE model by incorporating a novel causal disentanglement layer into VGAE, and it presents a significant advancement in causal inference methodology to achieve concept-free causal disentanglement, thereby enhancing our ability to analyze complex systems with minimal prior assumptions \cite{Feng2023ConceptfreeCD}.
The authors proposed a CaDeT approach, which significantly advances trajectory prediction for autonomous driving by incorporating causal reasoning into the learning process. The model enhances its predictive reliability in challenging and dynamic conditions by disentangling causal features and spurious features and utilizing a targeted intervention mechanism \cite{Pourkeshavarz2024CaDeTAC}.

\section{Problem Statement}
\begin{figure}[H]
	\centering
	\includegraphics[width=1.0\linewidth]{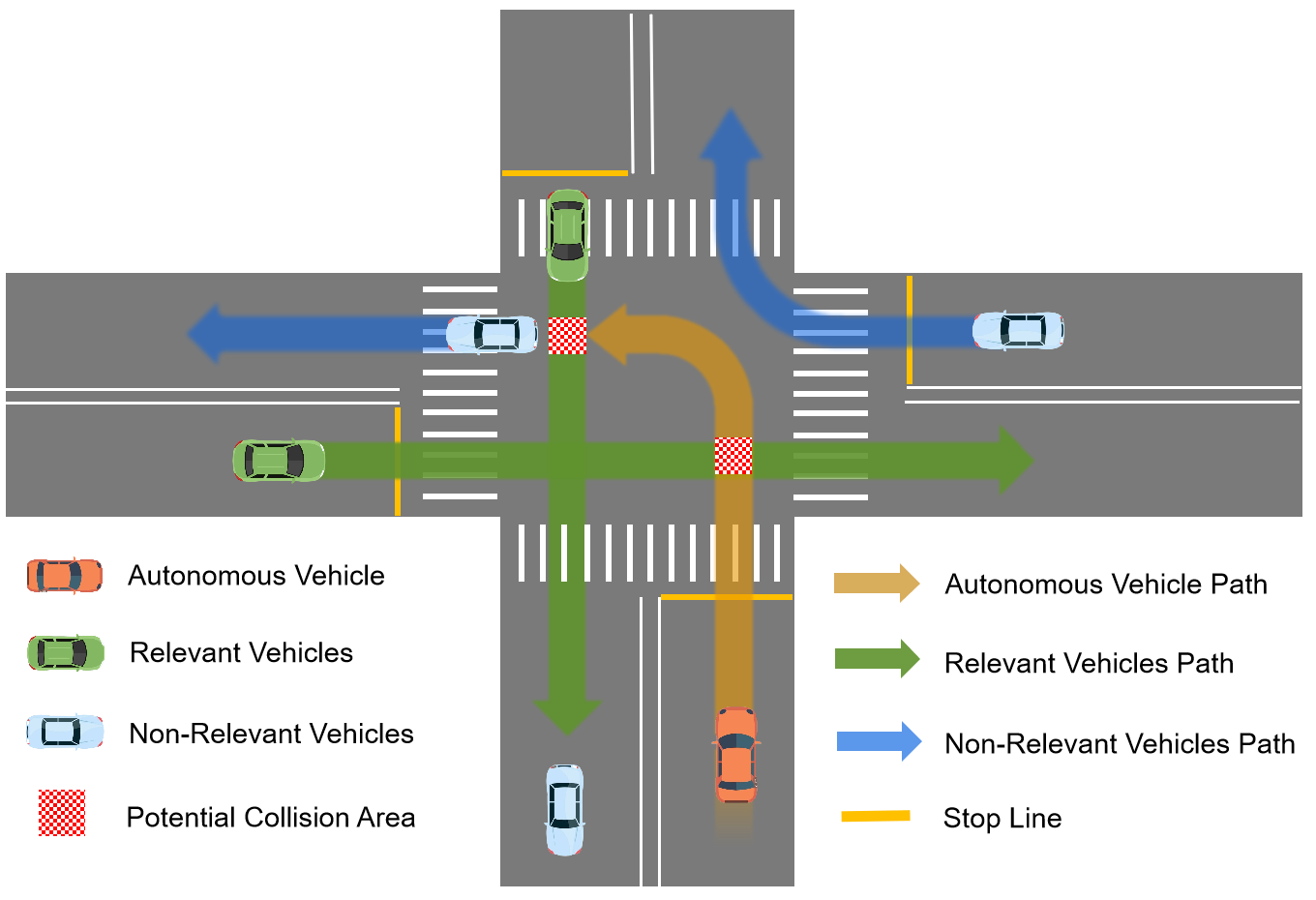}
	\caption{an unsignalized intersection scenario.}
	\label{fig:01}
\end{figure}
This study addresses the decision-making issue in autonomous driving at an unsignalized intersection, as depicted in Figure 1. The scenario involves a single autonomous ego vehicle (AEV) interacting with surrounding HVs. We demonstrate that the sequential decision-making process of the AEV can be effectively modeled as a Markov Decision Making (MDP). MDP stands as a pivotal mathematical construct for modeling complex decision-making scenarios. This framework performs well in environments where outcomes are influenced by both stochastic elements and deliberate actions taken by an agent, and it is widely employed in the field of autonomous driving to model various sequential decision-making problems. Furthermore, MDP is typically defined as a four-element tuple \(\langle \mathcal{S}_{av}, \mathcal{A}_{av}, \mathcal{R}_{av}, \mathcal{P}_{av} \rangle \). At each time step t, an autonomous vehicle interacts with their surrounding traffic environment and takes an action \(a_t\) based on the current state \(s_t\). Its action leads to a transition to the next state \(s_{t+1}\) according to transition probabilities \(\mathcal{P}_{av}\)\((s_{t+1}|s_{t},a_{t})\), resulting in a reward \(r_t\). \\
In our study, the number of HVs is predefined, with each appearing at random starting positions and assigned different destinations. In contrast, the AEV has fixed starting and destination points. We model the scenario as a graph, with nodes representing vehicles and edges capturing their pairwise interactions.
\begin{figure*}[ht]
	\centering
	\includegraphics[width=0.92\linewidth]{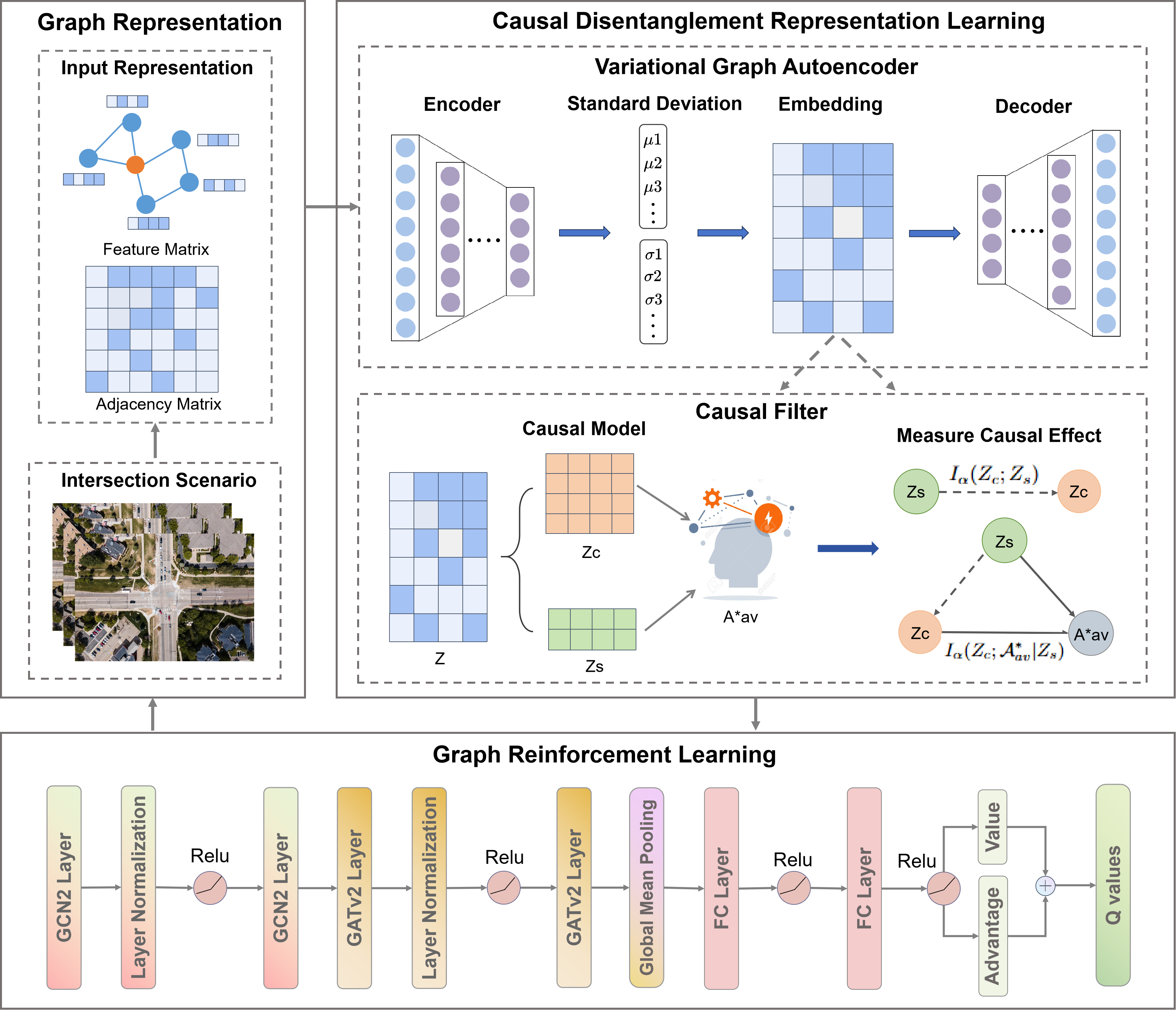}
	\caption{The framework of CGRL algorithm.}
	\label{fig:01}
\end{figure*}

\section{Causality-inspired Graph Reinforcement Learning}
In this section, we propose a novel multi-agent decision-making framework, CGRL, as illustrated in Figure 2. CGRL integrates CDRL with GRL to model the decision-making process of the AEV in multi-agent interactions at an unsignalized intersection scenario. Specifically, we leverage CDRL within VGAE to extract causal features from complex graph-structured data. These features are then input into the GRL algorithm.
\subsection{Reinforcement Learning Implementation}
In our experimental scenario, we primarily utilize an undirected and unweighted graph \(G = (\mathcal{V}, \mathcal{E})\) to model the interactions among vehicles, and in this graph, each vehicle is represented as a node, while the interaction between each pair of nodes is represented by an undirected edge. Here, \(\mathcal{V} = \{v_i, ...v_N\}\) represents the set of vehicles present in the current state. \(\mathcal{E}\) is the set of edges between vertices in the graph. An undirected edge is present between two vehicles if they are within a certain predefined distance from each other, specifically when the relative x-distance is less than 10 meters and the relative y-distance is less than 30 meters.\\
The state of the AEV \(\mathcal{S}_{av} = [F, A]\) consists of the feature matrix \(F\in\mathbb{R}^{N \times d}\) and the adjacency matrix \(A \in \mathbb{R}^{N \times N}\). \(F\) captures the attributes of each surrounding vehicle, while \(A\) represents the interactions between vehicles within a predefined distance of each other.
\subsubsection{Feature Matrix}
The feature matrix is composed of the attribute of each vehicle in the simulation, and the rows of \(F\) represent the number of vehicles, while the columns correspond to the feature dimensions for each vehicle. To distinguish the AEV from other vehicles, the first row of the features matrix is designated to represent the features of the AEV.
\begin{equation}
    F = [e^i, x^i, y^i, v_x^i, v_y^i, cos_h^i, sin_h^i]
\end{equation}
Where \(e^i\) denotes the presence or absence of the \(i^{th}\) vehicle, being set to 1 if the vehicle exists and 0 otherwise. \(x^i\) denotes the position of the \(i^{th}\) vehicle on the x-axis; \(y^i\) denotes the position of the \(i^{th}\) vehicle on the y-axis. \(v_x^i\) denotes the \(i^{th}\) velocity component in the x-direction. \(v_y^i\) denotes the \(i^{th}\) velocity component in the y-direction; $(\cos_h^i, \sin_h^i)$ denotes the trigonometric components of the \(i^{th}\) vehicle's heading angle, where $\cos_h^i$ and $\sin_h^i$ correspond to the cosine and sine of its heading angle, respectively.

\subsubsection{Adjacency Matrix}
The adjacency matrix is a square matrix that represents vehicle interactions. Each element \( e_{ij} \) in the matrix indicates whether a pair of vehicles \( i \) and \( j \) is adjacent. If vehicles \( i \) and \( j \) are within the specified range and interacting, \( e_{ij} = 1 \); otherwise, if there is no interaction or proximity, \( e_{ij} = 0 \).
\begin{equation}
A =
\begin{bmatrix}
e_{11} & e_{12} & \cdots & \cdots & e_{1n} \\
e_{21} & e_{22} & \cdots & \cdots & e_{2n} \\
\vdots & \vdots & \ddots & & \vdots \\
& & & e_{ij} & \\
\vdots & \vdots & & & \vdots \\
e_{n1} & e_{n2} & \cdots & \cdots & e_{nn}
\end{bmatrix}
\end{equation}

\subsubsection{Action Space (\(\mathcal{A}_{av}\))}
During interactions, the AEV operates within a discrete action space defined in our scenario. This action space allows it to perform three primary operations: accelerate, decelerate, or maintain a constant speed, corresponding to specific adjustments in its throttle and brake controls. At each time step, the AEV can choose one action from this space to navigate the intersection effectively.
\begin{equation}
    \mathcal{A}_{av}=\{constant,accelerated,decelerated\}
\end{equation}

\subsubsection{Reward Function (\(\mathcal{R}_{av}\))}
Each state-action pair is associated with a numerical reward signal. This reward \(\mathcal{R}_{av}\) represents the immediate benefit or cost of taking a specific action in a particular state. The AEV is designed with the goal of maximizing long-term cumulative rewards.
\begin{align}
     \mathcal {R}_{av} & = \omega_{t}^{or}*r_{t}^{or} * (\omega_{t}^{c}*r_{t}^{c} + \omega_{t}^{hs}*r_{t}^{hs} + \omega_{t}^{tc}*r_{t}^{tc})
\end{align}
where \(r_{t}^{c}\), \(r_{t}^{hs}\), \(r_{t}^{or}\), and \(r_{t}^{tc}\) represent a collision penalty term, a high speed reward term, a staying on road reward term, and a task completion term, respectively. The weights \(\omega\) correspond to their respective rewards, where \(\omega_{t}^{c}\), \(\omega_{t}^{hs}\), \(\omega_{t}^{or}\) and \(\omega_{t}^{tc}\) are 1 predefined as respectively.

\begin{itemize}
\item The component \(r_{t}^{c}\) is a reward signal related to collisions, indicating whether the vehicle has collided with another object or vehicle, This reward is penalizing to discourage the agent from engaging in unsafe behaviors that lead to collisions. Therefore, it is a critical component for teaching the AEV to drive safely and avoid accidents.
\begin{equation}
    r_{t}^{c}=\begin{cases}-2,&\text{if collision}\\0,&\text{otherwise.}\end{cases}
\end{equation}

\item The component \(r_{t}^{hs}\) is a reward signal that encourages the vehicle to maintain a high speed within a desired range. Moreover, this reward incentivizes the AEV to drive efficiently and reach its destination quickly, while still adhering to speed limits or safety constraints.
\begin{equation}
\begin{aligned}
    r_{t}^{hs} & = y_0 + \frac{(v - x_0) \cdot (y_1 - y_0)}{x_1 - x_0}
\end{aligned}
\end{equation}
Where \(v\) refers to the current velocity where the AEV has reached, 
The speed of the vehicle can be mapped from [\(x_0\), \(x_1\)] to [\(y_0\), \(y_1\)], which is then used to calculate the high-speed reward. 

\item The component \(r_{t}^{or}\) is a binary reward signal that indicates whether the vehicle is on the road. This reward encourages the agent to stay on the road, which is a fundamental requirement for safe and effective driving. This reward is crucial for ensuring that the AEV learns to follow the road and avoid dangerous situations.

\begin{equation}
    r_{t}^{or} =
\begin{cases} 
1, & \text{if on road} \\ 
0, & \text{otherwise.}
\end{cases}
\end{equation}

\item The component \(r_{t}^{tc}\) is the reward to encourage efficient task completion, when the vehicle successfully arrives the target location. As such, It serves as a strong incentive for the AEV to complete its mission successfully.
\begin{equation}
    r_{t}^{tc} =
\begin{cases} 
1, & \text{if arrive the destination} \\ 
0, & \text{otherwise.}
\end{cases}
\end{equation}
\end{itemize}

\subsection{Graph Neural Network Module}
We integrate the GAT with the GCN to process graph-structured data and use the dueling network as the GRL policy network. The absence of the GCN impairs the policy network’s ability to capture the chain reactions inherent in complex multi-agent interaction scenarios. Likewise, without the GAT, the network is unable to effectively prioritize the relative influence of each vehicle’s interactions.\\
Firstly, we mainly use two-layer GCN GCN2, which apply convolutional operations on graph-structure data. The equation is defined as:
\begin{equation}
\mathbf{x}^{(l+1)}=\left((1-\alpha)\Tilde{A}\mathbf{x}^{(l)}+\alpha\mathbf{x}^{(0)}\right)((1-\beta_l)\mathbf{I}+\beta_l\mathbf{W})
\end{equation}
Where \(\Tilde{A} = \mathbf{D}^{-1/2}\mathbf{\hat{A}}\mathbf{D}^{-1/2}\) is the symmetrically normalized adjacency matrix, \(\mathbf{\hat{A}} = \mathbf{A}+\mathbf{I}\) is the adjacency matrix with self-loops, \(\mathbf{D_{ii}}=\sum_{j} \mathbf{\hat{A}_{ij}}\) is diagonal degree matrix. \(\mathbf{x}^{(0)}\) is the initial node features, \(\mathbf{x}^{(l)}\) is the feature of layer \(l\). \(\alpha\) is the fraction of initial node features, and \(\beta_l\) is the hyperparameter to tune the strength of identity mapping. It is defined by \(\beta_l = \log(\frac{\lambda}{l} + 1) \approx \frac{\lambda}{l}\), where \(\lambda\) is a tunable hyperparameter. \(\beta\) ensures that the decay of the weight matrix progressively and adaptively intensifies as the network depth increases.
\\
After the GCN2 layers, input processed data into two-layer GATv2 \cite{Brody2021HowAA}, which perform attention-based aggregation of node features.
\begin{equation}
\alpha_{ij} = \frac{\exp(\mathbf{a}^\top \sigma (\left( \mathbf{W} \left[ \mathbf{x}_i \, \| \, \mathbf{x}_j \right] \right))}{\sum_{k \in \mathcal{N}(i)} \exp(\mathbf{a}^\top \sigma \left( \mathbf{W} \left[ \mathbf{x}_i \, \| \, \mathbf{x}_k \right] \right)})
\end{equation}
where \( || \) is the concatenation operation; \(\mathbf{W}\) is a learnable weight vector; \( \sigma(\cdot) \) is the LeakyReLU activation function. \( \alpha_{ij} \) quantifies the significance of node \( j \)’s features for node \( i \).\\ Use the attention weights \(\alpha_{ij}\) to aggregate the features of neighboring nodes and update the representation of node i
\begin{equation}
\mathbf{x}'_i = \sigma \left( \sum_{j \in \mathcal{N}(i)} \alpha_{ij} \mathbf{W} \mathbf{x}_j \right)
\end{equation}
where \(\sigma(\cdot)\) is applied component-wise, The set \( N_i \) includes node \( i \) along with its neighboring nodes.\\
In order to enhance the expressive power of the model, GATv2 utilizes a multi-head attention mechanism, concatenating the results from M attention heads.
\begin{equation}
\mathbf{x}''_i = \big\|_{m=1}^M \sigma \left( \sum_{j \in \mathcal{N}(i)} \alpha_{ij}^{(m)} \mathbf{W}^{(m)} \mathbf{x}_j \right)
\end{equation}
Where \(\alpha_{ij}^{(m)}\) is the weight of the \(m\)-th attention head. 
\(W^{(m)}\) is the weight matrix of the \(m\)-th attention head.  
\(\parallel\) Represents the concatenation of outputs from different heads.\\
Capturing global features that aggregate information from all nodes is essential. After processing through the GATv2 layers, global average pooling is applied to compute the mean feature representation across all nodes in the graph.
\begin{equation}
    h = \frac{1}{N} \sum_{i=1}^{N} \mathbf{x}''_i
\end{equation}
Finally, the global features are mapped to the hidden layer through two fully connected layers, and the transformed global features are used as input for the dueling
layer to calculate the Q-value. In the dueling architecture, the state value estimation and advantage calculation are processed through separate pathways, both utilizing shared features extracted by a common backbone network. The Q-value formula is given by:
\begin{equation}
    Q(s,a) = V'(s) + A'(s,a) - \frac{1}{|A_{av}|} \sum_{a'} A'(s, a')
\end{equation}
Where \(V'(s)\) is the state value computed by the value network. \(A'(s,a)\) is the advantage for action a at state s. \(\frac{1}{|A_{av}|} \sum_{a'} A'(s,a') \) normalizes the advantages across all possible actions to ensure that the advantage reflects the relative improvement of action a over others. \(|A_{av}|\) is the number of possible actions. 

\subsection{Deep Reinforcement Learning Module}
The deep reinforcement learning component of our GRL architecture is implemented using the Dueling Double Deep Q-Network (D3QN) algorithm \cite{pmlr-v48-wangf16}, combining it with GNN module to model the decision-making process in autonomous driving. Firstly, D3QN can enhance the original DQN by integrating the principles of Double Q-learning, reducing the overestimation bias inherent in DQN while maintaining its ability to handle reinforcement learning problems in high-dimensional state and action spaces. Secondly, similar to DQN, D3QN uses a deep neural network \( Q(s, a; \theta) \) to approximate the state-action values, where \( \theta \) is the parameter of the network.
D3QN modifies the temporal difference learning process by decoupling action selection and evaluation, which helps mitigate overestimation bias. Originally, the optimal action-value function in D3QN is updated using the following rule:
\begin{equation}
    Q(s_t, a_t) \leftarrow Q(s_t, a_t) + \alpha \cdot \left[ r_t + \gamma Q_{\text{target}}(s_{t+1}, a_{\text{max}}) - Q(s_t, a_t) \right]
\end{equation}
where \( a_{\text{max}} = \arg\max_a Q_{\text{online}}(s_{t+1}, a; \theta) \) is determined by the online network, and the value is evaluated using the target network \( Q_{\text{target}}(s_{t+1}, a_{\text{max}}; \theta^-) \).

The true target values of D3QN in our algorithm can be expressed as:
\begin{equation}
    y = \mathcal{R}_{av} + \gamma Q_{\text{target}}(\mathcal{S'}_{av}, \arg\max_{a \in \mathcal{A}_{av}(\mathcal{S'}_{av})} Q_{\text{online}}(\mathcal{S'}_{av}, a; \theta); \theta^-)
\end{equation}
Here, \( \theta \) represents the trainable parameters of the online network, and \( \theta^- \) corresponds to the fixed parameters of the target network.

The loss function is designed to minimize the temporal difference error while leveraging the decoupled action selection and evaluation to improve stability:
\begin{equation}
    L = \mathbb{E}\left[\left(y - Q_{\text{online}}(\mathcal{S}_{av}, \mathcal{A}_{av}; \theta)\right)^2\right]
\end{equation}

Gradient descent is used to iteratively update the parameters \( \theta \) of the online network in the direction that decreases the loss function. The loss function gradients with respect to the network parameters are derived as follows:
\begin{equation}
    \nabla_\theta L = \mathbb{E}\left[\left(y - Q_{\text{online}}(\mathcal{S}_{av}, \mathcal{A}_{av}; \theta)\right) \nabla_\theta Q_{\text{online}}(\mathcal{S}_{av}, \mathcal{A}_{av}; \theta)\right]
\end{equation}

\subsection{Causal Disentangled Representation Learning with VGAE}
In this subsection, we employ the CDRL method within the VGAE framework. Unlike traditional causal discovery methods, CDRL is an emerging field of research that seeks to address the challenge of extracting causal features from complex, high-dimensional data.
\subsubsection{Causal Graphical Model}
We utilize a directed acyclic graph as a Causal Graphical Model to describe the causal relationships among multiple variables and construct a Structural Causal Model (SCM). This includes graph data G, two isolated features derived from the hidden space in VGAE: causal features \(Z_c\) and spurious features \(Z_s\), as well as the optimal decision-making of the autonomous vehicle \(\mathcal{A}^{*}_{av}\), where the cause-effect relationship is listed as \(Z_c\) ← \(G\) → \(Z_s\) → \(\mathcal{A}^{*}_{av}\). Below is a detailed explanation of the components of this causal graphical model:
\begin{itemize}
    \item \(Z_c\) ← \(G\) → \(Z_s\). The causal features \(Z_c\) serves as a key descriptor of the graph data \(G\), accurately reflecting its intrinsic properties and exerting a direct causal influence on the optimal decision-making of the autonomous vehicle \(\mathcal{A}^{*}_{av}\). The spurious feature \(Z_s\), typically caused by data biases or noise and does not have a direct causal relationship with \(\mathcal{A}^{*}_{av}\), but this influence is often spurious in nature. Since \(Z_c\) and \(Z_s\) naturally coexist in graph data G, these causal effects are established.
    \item \(Z_s\) → \(\mathcal{A}^{*}_{av}\). In this context, the spurious feature \(Z_s\) acts as a confounder between \(Z_c\) and \(\mathcal{A}^{*}_{av}\), because \(Z_s\) is causally connected to both \(Z_c\) and \(\mathcal{A}^{*}_{av}\), potentially creating a spurious association between \(Z_c\) and \(\mathcal{A}^{*}_{av}\) if \(Z_s\) is not properly accounted for.
\end{itemize}

\subsubsection{Graph-based Generative Model}
After constructing a Causal Graphical Model, we use VGAE to map above-mentioned high-dimensional graph-structured data, including the feature and adjacency matrices, into a low-dimensional latent space to generate low-dimensional feature matrix \(Z \in \mathbb{R}^{N \times (L)}\). Additionally, applying CDRL within VGAE to extract the causal features, which serve as inputs for the GRL model. Our VGAE model comprises two main components: an encoder and a decoder. \\
We employ a two-layer GCN as the encoder for the VGAE. The encoder processes the graph data and outputs the latent variable \(Z\), which captures the underlying representations of the graph structure and feature features. The first GCN layer generates a lower-dimensional feature matrix. It is defined as:
\begin{equation}
\bar{F} = \text{GCN}(F, A) = \text{ReLU}(\Tilde{A} F W_0)
\end{equation}
Where \(\Tilde{A} = D^{-\frac{1}{2}} A D^{-\frac{1}{2}}\) is the symmetrically normalized adjacency matrix, while \(D\) is degree matrix of \(A\). \(\bar{F}\) is feature Outputs after applying the first GCN layer. \(W_0\) is Learnable weight matrix of the first GCN layer, mapping input features to a lower-dimensional space.\\
The second GCN layer generates \(\mu\) and \(\sigma\), as in the following equations:
\begin{equation}
\begin{aligned}
\mu &= GCN_\mu(F, A) = \Tilde{A} \bar{F} W_{1} \\
\log(\sigma^2) &= GCN_\sigma(F, A) = \Tilde{A} \bar{F} W_1
\end{aligned}
\end{equation}
Where \(\mu\) is the mean vector matrix, \(\sigma\) is the variance matrix, \(\log \sigma\) and \(\mu\) share the weight \(W_{1}\). Then we can obtain latent vector Z.
\begin{equation}
\begin{aligned}
q(Z \mid F, A) &= \prod_{i=1}^N q(z_i \mid F, A) \\
q(z_i \mid F, A) &= \mathcal{N}(z_i \mid \mu_i, \text{diag}(\sigma_i^2))
\end{aligned}
\end{equation}
Where \(\mathbf{Z} \in \mathbb{R}^{N \times L}\) be the latent representation, with \(N\) denoting the number of vehicles and \(L\) specifying the dimensionality of the latent representation. Additionally, the latent representation for the \(i_{th}\) vehicle \(v_i\) is denoted as \(\mathbf{z}_i\).\\
In VGAE, we can utilize a multi-layer perception paired with an inner product decoder. Its generative model can be formulated as:
\begin{equation}
\begin{aligned}
p(A \mid Z) &= \prod_{i=1}^N \prod_{j=1}^N p(a_{ij} \mid z_i, z_j) \\
p(a_{ij} &= 1 \mid z_i, z_j) = \sigma(z_i^\top z_j)
\end{aligned}
\end{equation}
The overall objective of VGAE is including optimizing the evidence lower bound (ELBO) of VAE while simultaneously minimizing the total correlation, which can be formulated as
\begin{equation}
    \mathcal{L}_{\mathbf{VGAE}}=E_{q(Z|F,A)}[logp(A|Z)]-KL[q(Z|F,A)||p(Z)]
\end{equation}
Where \(KL\) is the Kullback-Leibler divergence.
\subsubsection{Causal Filter}
After generate low-dimensional feature matrix \(Z\), the challenge is how to identify and extract causal features. To address this challenge, we employ state-of-the-art information-theoretic frameworks to quantify and optimize causal relationships within the system \cite{Ay2008InformationFI}, and information-theoretics have been utilized in GNN \cite{Zheng2023CIGNNAG, Lin2022OrphicXAC}. To find causal features \(Z_c\), we first ensure that \(Z_c\) and \(Z_s\) are independent, Subsequently, causal intervention techniques are applied to maximize the causal influence between \(Z_c\) and \(\mathcal{A}^{*}_{av}\). This approach effectively removes the spurious features, uncovering the true causal relationship and extracting the natural causal features \(Z_c\).\\
Mutual information (MI) \(I\left(Z_c; Z_s\right)\) is used to ensure whether the causal features \(Z_c\) and the spurious feature features \(Z_s\) are independent, as MI measures the statistical dependence between two random variables between \(Z_c\) and \(Z_s\), such as \(I\left(Z_c; Z_s\right) = 0\) means \(Z_c\) is independent of \(Z_s\). The estimation method of MI is originally defined as:
\begin{equation}
    I(Z_c;Z_s)=\mathbb{E}_{P(Z_c,Z_s)}\left[\log\frac{P(Z_c,Z_s)}{P(Z_c)P(Z_s)}\right]
\end{equation}
However, the method based on probability distribution requires knowledge of the joint distribution P(\(Z_c\),\(Z_s\)) and marginal distributions P(\(Z_c\)) and P(\(Z_s\)), which are often unavailable in practice, and direct estimation of MI is difficult for high-dimensional variables, where exact probability distributions are unknown. \\
Other methods use entropic graph techniques have been developed that eliminate the need for explicit distribution estimation. include the k-nearest-neighbour \cite{Pl2010EstimationOR} and the MI Neural Estimators\cite{Belghazi2018MINEMI}. However, the MI Neural Estimator presents several challenges. The first one is instability during training, joint optimization of the neural network can lead to issues like unstable convergence or poor local minima, and the logarithmic terms and exponentiation can result in numerical instability. The second one is negative MI values, In some cases, especially when the model is undertrained, the estimator may output negative MI values, which are not valid since MI is always non-negative. The third one is high computational cost, training the neural network can be computationally expensive, particularly for large datasets or high-dimensional inputs. For the k-nearest-neighbour estimator, its non-differentiability prevents the use of traditional optimization algorithms, such as gradient descent, which rely on the differentiability of the objective function for parameter adjustment.\\
Therefore, Sánchez Giraldo et al. \cite{SanchezGiraldo2012MeasuresOE} introduced a matrix-based functional to compute Rényi \(\alpha\)-order entropy, characterizing both entropy and MI through the normalized eigenvalues of the Gram matrix. This Gram matrix, a Hermitian operator, is derived by mapping the data into a reproducing kernel Hilbert space. This approach is more effective for MI estimation, as matrix-based methods are naturally suited for high-dimensional data and do not rely on neural network optimization. This avoids issues such as unstable training, sensitivity to hyperparameters, and ensures reproducible results. Additionally, it directly computes MI from the eigenspectrum of similarity matrices, bypassing the need for joint and marginal distribution estimation. Therefore, this method is especially useful in scenarios where stability and computational efficiency are crucial. \(I\left(Z_c; Z_s\right)\) can be represented as:
\begin{equation}
    I_\alpha(Z_c; Z_s)=S_\alpha(Z_c)+S_\alpha(Z_s)-S_\alpha(Z_c, Z_s)
\end{equation}
The \(\alpha\)-order Rényi entropy for one matrix \(Z_c\) or \(Z_s\) is defined as:
\begin{equation}
\begin{aligned}
S_\alpha(Z_c) &= \frac{1}{1-\alpha}\log_2\mathrm{tr}((Z_c)^\alpha) \\
&= \frac{1}{1-\alpha}\log_2\left(\sum_{i=1}^n\lambda_i(Z_c)^\alpha\right)
\end{aligned}
\end{equation}
where \(\lambda_i(Z_c)\) are the eigenvalues of the matrix \(Z_c\), and \(\mathrm{tr}((Z_c)^\alpha)\) is the trace of \((Z_c)^\alpha\). The parameter \(\alpha\) is used to define different orders of entropy.\\
For two matrices \(Z_c\) and \(Z_s\), the joint Rényi entropy can be defined similarly:
\begin{equation}
\begin{aligned}
S_\alpha(Z_c, Z_s) &= \frac{1}{1-\alpha}\log_2\mathrm{tr}(D^\alpha) \\
\end{aligned}
\end{equation}
Where \(D = Z_c \otimes Z_s\) denotes the Kronecker product of matrices \(Z_c\) and \(Z_s\).\\
Due to the incorrect decision-making \(\mathcal{A}^{*}_{av}\) based on spurious features \(Z_s\) instead of causal features \(Z_c\), it is crucial to eliminate the backdoor path.
To identify direct causal connections between \(Z_c\) and \(\mathcal{A}^{*}_{av}\). we can apply the do-calculus on the variable \(Z_s\) to block the backdoor path. Accordingly, we propose an intervention mechanism that leverages a conditional MI method to quantify causal influence and identify the causal relationship by intervening on \(Z_s\).
\begin{equation}
    I\left(Z_c\to \mathcal{A}^{*}_{av}\right|do\left(Z_s\right)) = I\left(Z_c;\mathcal{A}^{*}_{av}|Z_s\right)
\end{equation}
We also use the Matrix-based Rényi entropy method to directly estimate the conditional MI.
\begin{equation}
\begin{aligned}
I_\alpha(Z_c;\mathcal{A}^{*}_{av}|Z_s) &= S_\alpha(Z_c|Z_s)+S_\alpha(\mathcal{A}^{*}_{av}|Z_s)-S_\alpha(Z_c,\mathcal{A}^{*}_{av}|Z_s) \\
&= S_\alpha(Z_c,Z_s) + S_\alpha(\mathcal{A}^{*}_{av}|Z_s)\\
&- S_\alpha(Z_s) - S_\alpha(Z_c,\mathcal{A}^{*}_{av},Z_s)
\end{aligned}
\end{equation}
The joint Rényi entropy for three variables can be defined as:
\begin{equation}
    S_\alpha(Z_c,Z_s,\mathcal{A}^{*}_{av})=\frac{1}{1-\alpha}\log_2\mathrm{tr}((K_{3})^\alpha)
\end{equation}
Where \(K_{3}\) where is the Gram matrix constructed from the joint distribution of \(Z_c\), \(Z_s\) and \(\mathcal{A}^{*}_{av}\).\\
This step involves using a decoder to map the extracted causal features to a causal adjacency matrix, which then serves as the state input for training the CGRL decision-making algorithm.
\begin{equation}
    A_c=\sigma(Z_cZ_c^T)
\end{equation}
\subsubsection{Objective function}
To obtain causal features \(Z_c\) from the latent feature matrix \(Z\), the overall loss function can be defined based on \cite{Lin2022OrphicXAC}:
\begin{equation}
    \min\begin{array}{c}- I_\alpha(Z_c;\mathcal{A}^{*}_{av}|Z_s)+I_\alpha(Z_c; Z_s)+\lambda_1\mathcal{L}_{\mathbf{VGAE}}+\lambda_2\frac{\|A_c\|_1}{\|A\|_1}\end{array}
\end{equation}
where \(\lambda_i\) \((i \in {\{1, 2\}})\) controls the associated regularizer terms.
The term \(- I_\alpha(Z_c;\mathcal{A}^{*}_{av}|Z_s)\) is minimized to maximize the causal influence between \(Z_c\) and \(\mathcal{A}^{*}_{av}\). This process involves removing the spurious influence of \(Z_s\) and ensuring that the causal pathway between \(Z_c\) and \(\mathcal{A}^{*}_{av}\) remains strong and direct, unaffected by any extraneous confounding factors.\\
Minimizing \(I_\alpha(Z_c; Z_s)\) is designed to ensure the independence between the causal features \(Z_c\) and the spurious feature \(Z_s\). This term is crucial because it forces the model to eliminate any potential correlation or indirect dependency between the two feature sets. \\
\(\mathcal{L}_{\mathbf{VGAE}}\) is the negative evidence lower bound (ELBO) loss term, and minimize it can help approximate the true posterior distribution of latent variables, improves the approximation of the posterior distribution of the latent variables, which leads to better representations of graph data.\\
Minimizing \(\frac{\|A_c\|_1}{\|A\|_1}\) is to minimize the number of nodes and edges necessary to represent the key information. 

\section{Experiment}
\subsection{Driving Scenario Setup}
To assess the performance of our CGRL approach, Our experiment employs the highway-env simulator \cite{highway-env} to create an unsignalized intersection scenario with 15 human-driven vehicles, which is an open-sourced simulation platform designed for developing and testing decision-making algorithms in autonomous driving systems. The experimental setup involves a four-way intersection, where each road measures 4 meters in width and extends 30 meters in length in all directions. The main road usually has a higher priority, allowing vehicles to pass without slowing down, while vehicles on the secondary road need to observe the main road traffic before entering the intersection and then proceed when it is safe. In the simulation, we employ intelligent driver model (IDM)  \cite{Treiber2000CongestedTS} as a fundamental traffic flow model for vehicle dynamics. Vehicles are modeled to follow this car-following behavior, adjusting their speed based on the relative distance to the preceding vehicle. This allows for the modeling of realistic interactions between vehicles at the unsignalized intersection. \\
\begin{table}[H]\tiny
  \centering
  {\fontsize{10pt}{10pt}\selectfont
  \renewcommand{\arraystretch}{1.3}
  \setlength{\tabcolsep}{15pt}
  \caption{IDM parameters used for training.}
  \label{table:hyperparameters}
  \begin{tabular}{@{\hspace{1em}} l c r @{\hspace{1em}}} \hline
  \textbf{Keyword} & \textbf{Value} & \textbf{Unit} \\ \hline
Maximum acceleration $a_\text{max}$ & 6              & m/s$^2$       \\ 
Acceleration argument $\delta$          & 4              & /             \\ 
Desired time gap $T$                    & 1.5            & s             \\
Minimum jam distance $s_0$                    & 5            & m             \\
Comfortable deceleration $b$ & -5             & m/s$^2$       \\  \hline
  \end{tabular}
  }
\end{table}
In terms of IDM model, it is a car-following model, simulates how vehicles adjust their speed based on the car ahead. It considers factors like desired velocity \(v_0\), the gap distance between the vehicle and the leading vehicle \(s\), the maximum acceleration of the vehicle \(a_{\text{max}}\), and the speed of the vehicle \(v\) to replicate realistic driving behaviors.
\begin{equation}
    a = a_{\text{max}} \left[1 - \left(\frac{v}{v_0}\right)^\delta - \left(\frac{s^*(v,\Delta v)}{s}\right)^2\right]
\end{equation}
where \(s^*(v,\Delta v)\) is the safety distance function, which depends on the current speed \(vT\) and the speed difference \(\Delta v\).
\begin{equation}
    s^*(v,\Delta v) = s_0 + vT + \frac{v\Delta v}{2\sqrt{a_{\text{max}}b}}
\end{equation}
Where \(s_0\) is the minimum jam distance to the vehicle ahead. \(vT\) represents the distance corresponding to the desired time gap. \(\Delta v\) represents the difference between the vehicle's speed and the speed of the vehicle ahead. b is the comfortable deceleration.
\subsection{Implementation details}
\subsubsection{Training details}
The proposed decision-making method is employed to train an optimal behavior policy for the autonomous vehicle. Specifically, a PyTorch-based decision-making framework is utilized to train the model over 1000 episodes. Each episode concludes once the autonomous vehicle successfully arrives at the target destination or collides with any other road user. Additionally, the reward values for each episode are recorded, making it convenient to conduct model testing. Simulation hyperparameters are summarized in TABLE II.
\begin{table}[H]\tiny
  \centering
  {\fontsize{10pt}{10pt}\selectfont
  \renewcommand{\arraystretch}{1.3}
  \setlength{\tabcolsep}{15pt}
  \caption{Hyperparameter used for training.}
  \label{table:hyperparameters}
  \begin{tabular}{@{\hspace{1em}} l c r @{\hspace{1em}}} \hline
  \textbf{Parameters} & \textbf{Symbols} & \textbf{Value} \\ \hline
Discounted factor     & \(\gamma\)                   & 0.95  \\ 
Replay memory size    & \( M_{\text{replay}} \)      & 100000    \\ 
Mini-batch size       & \( M_{\text{mini}} \)        & 64      \\ 
Learning rate & \( \eta \) & 0.0001 \\ 
Epsilon & \(\epsilon\) & 0.1 \\ 
Target update frequency & $N_{\text{update}}$ & 5000 \\ \hline
  \end{tabular}
  }
\end{table}
\subsubsection{Network Architecture}
\begin{figure*}[htbp]
\centering

\begin{minipage}{0.32\textwidth}
    \centering
    \includegraphics[width=\linewidth]{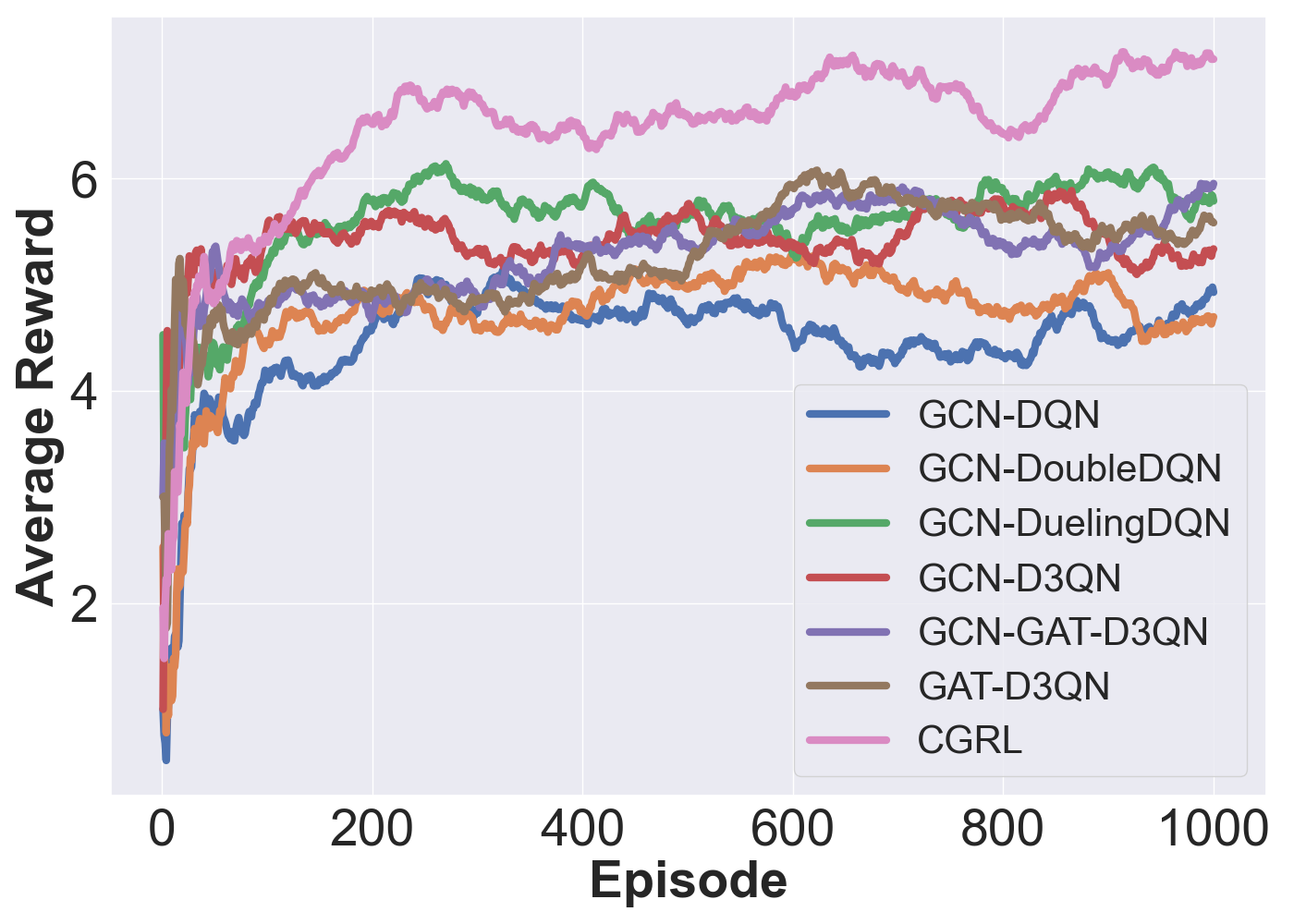}\\
    (a)
\end{minipage}
\begin{minipage}{0.32\textwidth}
    \centering
    \includegraphics[width=\linewidth]{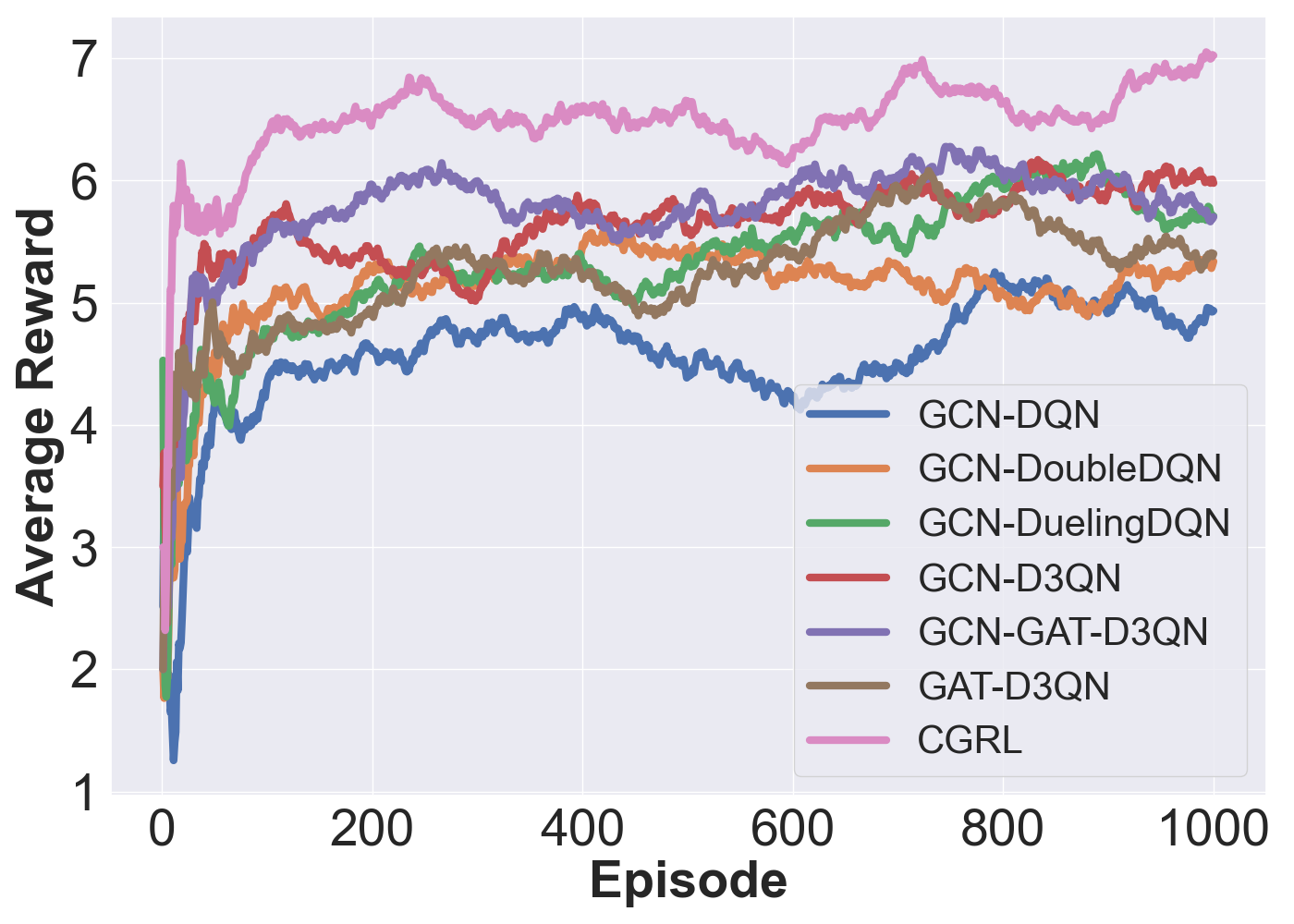}\\
    (b)
\end{minipage}
\begin{minipage}{0.32\textwidth}
    \centering
    \includegraphics[width=\linewidth]{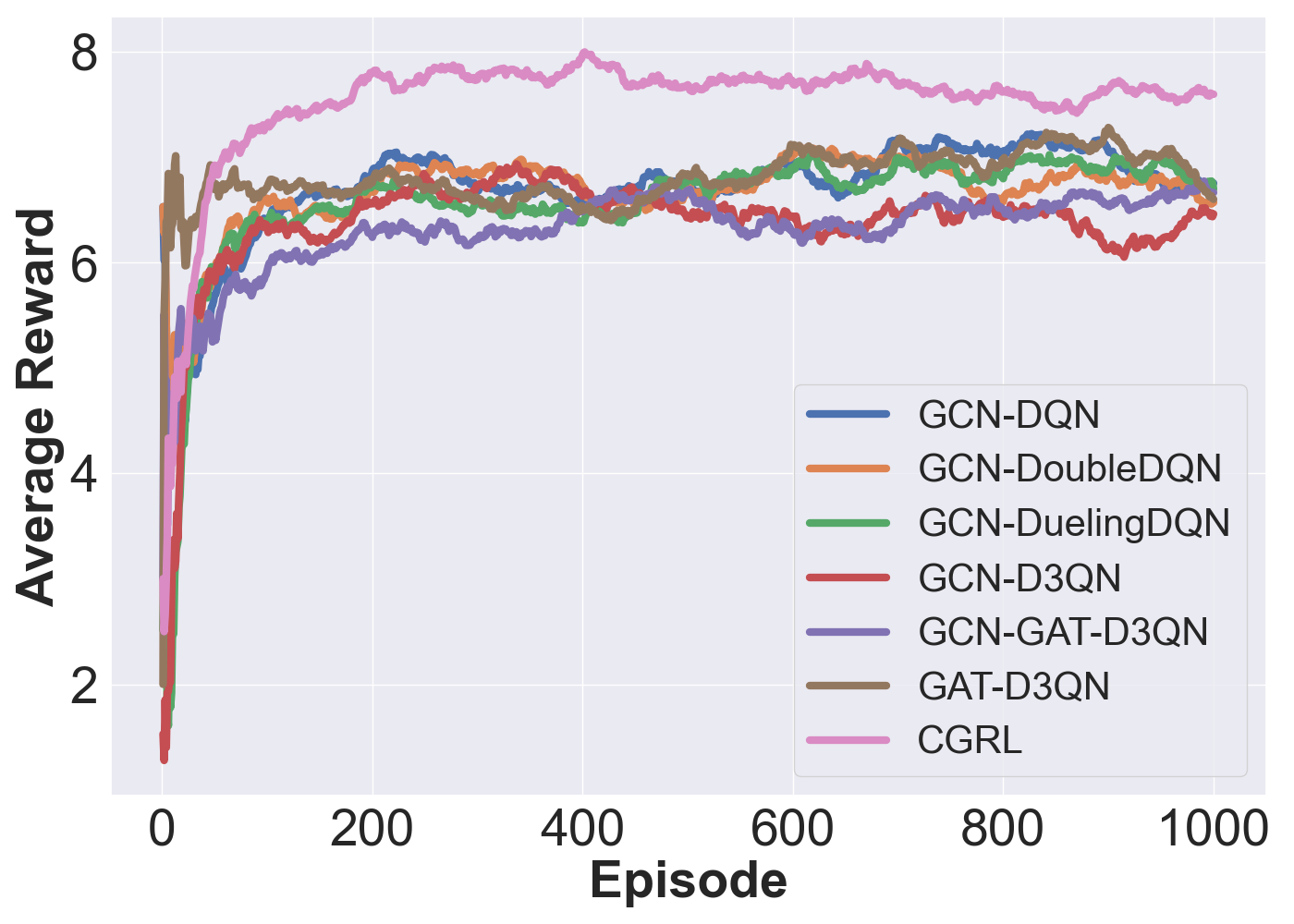}\\
    (c)
\end{minipage}

\vspace{1em}

\begin{minipage}{0.32\textwidth}
    \centering
    \includegraphics[width=\linewidth]{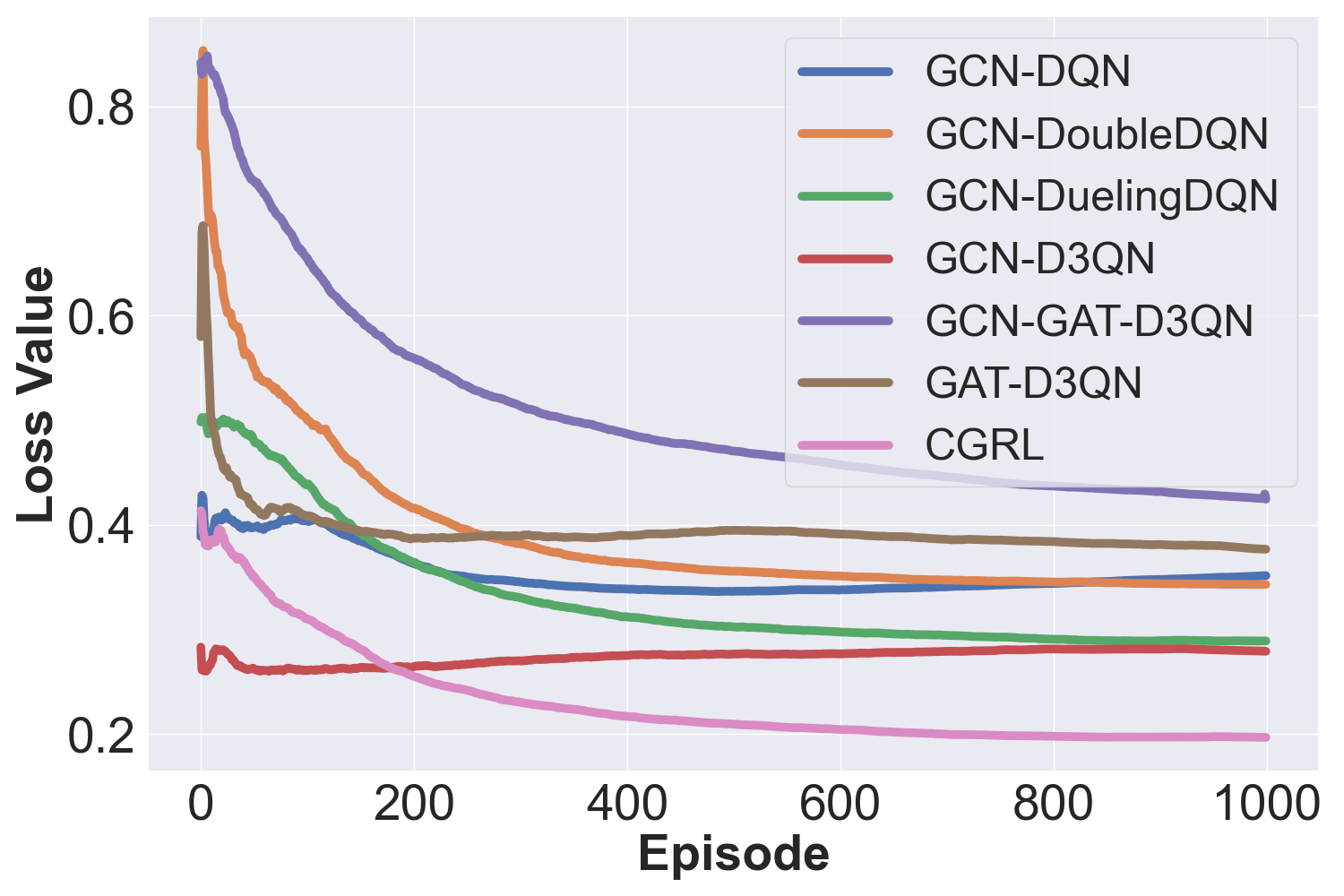}\\
    (d)
\end{minipage}
\begin{minipage}{0.32\textwidth}
    \centering
    \includegraphics[width=\linewidth]{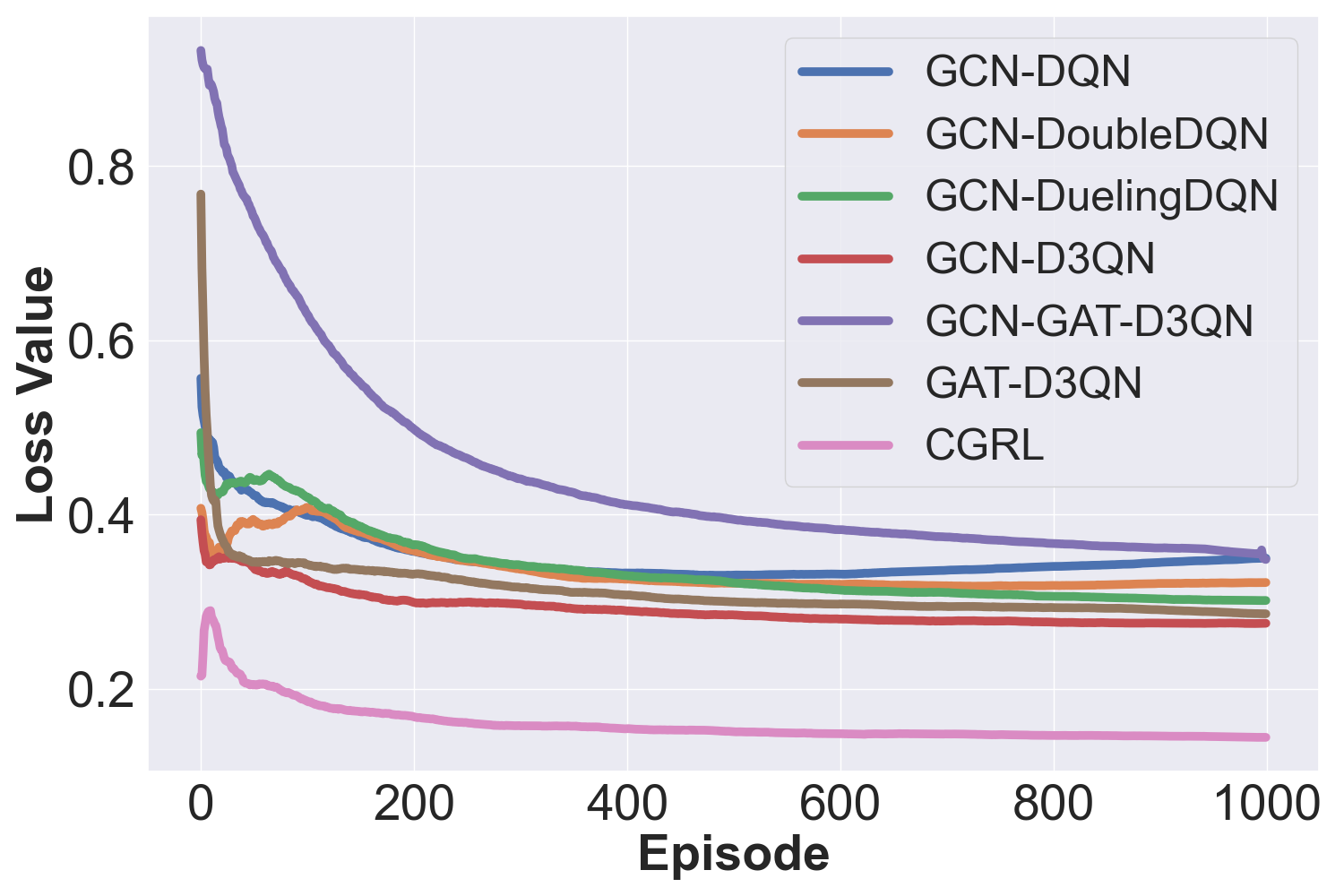}\\
    (e)
\end{minipage}
\begin{minipage}{0.32\textwidth}
    \centering
    \includegraphics[width=\linewidth]{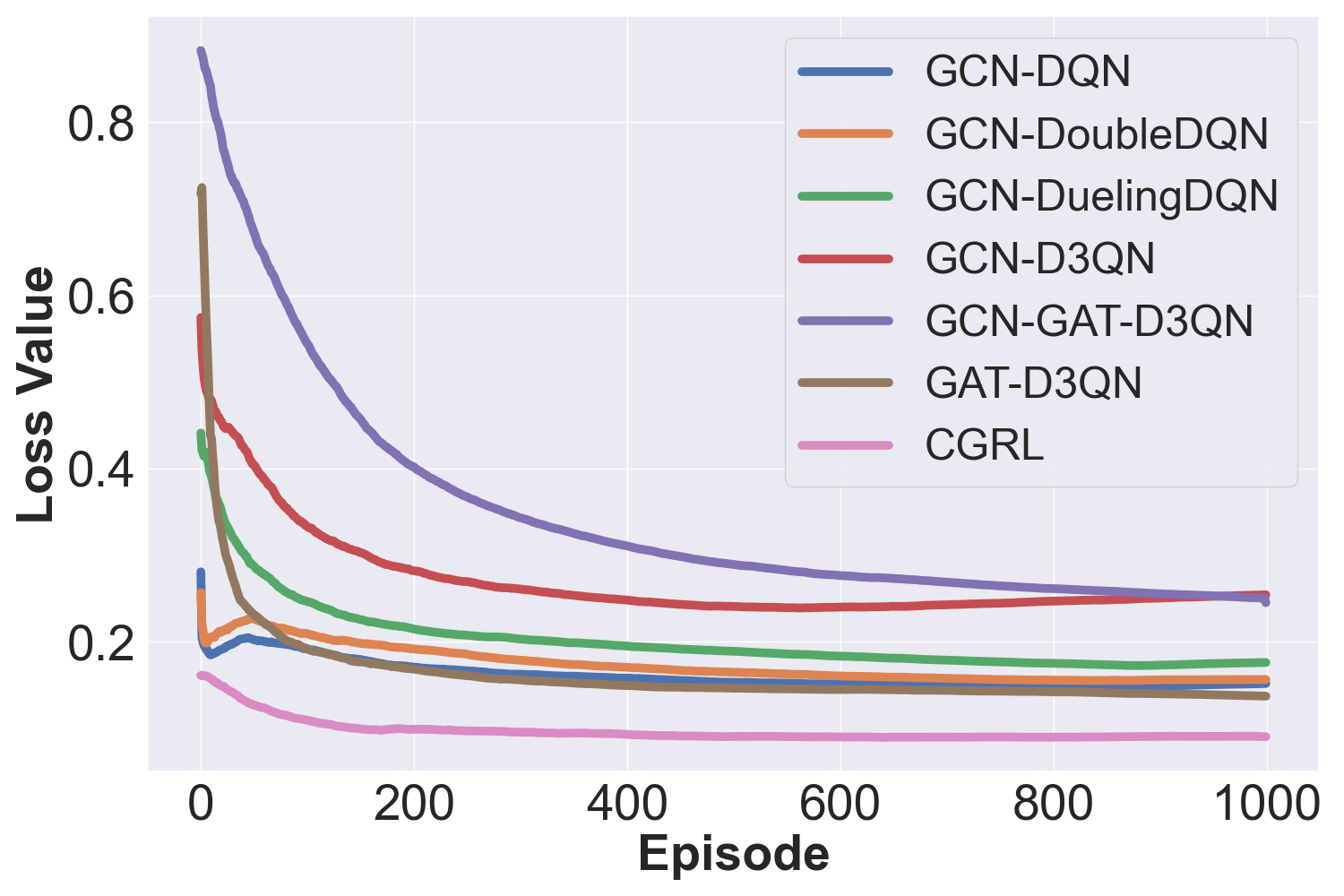}\\
    (f)
\end{minipage}

\vspace{1em}


\caption{Training Rewards and Loss Curves Across Three Driving Tasks: Left Turn, Straight, and Right Turn.
The reward and loss values for the left-turn task are shown in subfigures (a) and (d); for the straight-driving task in (b) and (e); and for the right-turn task in (c) and (f).}
\label{fig:training_summary}
\end{figure*}
The network architecture of our proposed CGRL algorithm is illustrated in Figure 2. First, the input node features pass through two GCN2 layers to extract local structural information, followed by two GATv2 layers to capture global attention relationships . The first two-layer GATv2 takes processed features as input and produces embeddings, which are further processed by the two-layer GATv2. ReLU activation and Layer Normalization is applied after the first GCN2 and the first GATv2 layer.\\
The outputs from the second GATv2 layer are aggregated through a mean pooling operation across all nodes, resulting in a single vector representation of the entire graph. This aggregated graph embedding is then passed through two fully connected layers to generate latent representations, with ReLU activation applied once again to enhance non-linearity.\\
The dueling network splits the latent representation into two streams: an advantage network and a value network. The advantage network maps the latent representation to action-specific advantage values, outputting a vector. The value network outputs a scalar value representing the overall state value. At last, a fully connected output layer computes the final Q-values.
\subsubsection{Performance Metrics}
To assess and validate the performance of our method, we conduct tests on each trained policy and record several key metrics, including collision rate, average velocity, and average reward.
\begin{itemize}
\item[$\bullet$] Collision rate (C.R.): The frequency of collisions between an autonomous vehicle and other surrounding vehicles during testing
\end{itemize}
\begin{itemize}
\item[$\bullet$] Average velocity (A.V.): The average driving speed of an autonomous vehicle during testing.
\end{itemize}
\begin{itemize}
\item[$\bullet$] Average Reward (A.R.): This metric represents the average accumulated rewards achieved by an autonomous vehicle during testing.
\end{itemize}
\subsubsection{Comparison Baselines}
GCN-DQN, GCN-Double-DQN, GCN-Dueling-DQN, GCN-D3QN \cite{Liu2022GraphCD}, GAT-D3QN \cite{Cai2021DQGATTS}, and GCN-GAT-D3QN \cite{Zhou2024ReasoningGR} are used as baseline methods to evaluate the effectiveness of integrating graph-based representations with various DQN variants. These baselines provide a basis for comparison against the proposed method.\\
GCN-DQN: The method combines the GCN with DQN to capture the relational structure of agents or entities in a graph-based environment, enabling more informed and coordinated decision-making.\\
GCN-Double-DQN: The method combines the GCN with Double DQN to enable more effective learning in multi-agent interaction environments.\\
GCN-Dueling-DQN: The method combines the GCN with Dueling DQN to improve decision-making in graph-structured environments by extracting relational features among agents and separating value and advantage estimations for more stable and efficient learning.\\
GCN-D3QN: This approach integrates GCN with D3QN to enhance decision-making in autonomous vehicle interaction scenarios, utilizing GNN to extract scenario features and the D3QN framework to generate driving behaviors.\\
GAT-D3QN: The method combines the GAT with D3QN to achieve improved autonomous vehicles decision-making in interactive scenarios. In this approach, it employs GAT to extract features from interactive scenarios, utilizing attention mechanisms to emphasize the significance of inter-vehicle relationships. These features are then fed into the D3QN framework, which generates optimal decisions for autonomous vehicles in dynamic and interactive environments.\\
GCN-GAT-D3QN: The method combines the GCN, GAT, and D3QN to capture both global structural information and fine-grained relational importance among agents, while improving the stability and efficiency of value-based reinforcement learning in graph-structured multi-agent environments.
\section{Results}

With the necessary setup completion, this section involves initiating the simulation and training loop to execute the specified number of episode. During this process, the CGRL algorithm's performance in the unsignalized intersection environment is assessed on three distinct driving tasks: left turn, straight traversal, and right turn. Episodic rewards are recorded to assess performance during training. Additionally, comparative tests are conducted to statistically analyze the performance of all of decision-making algorithms. The testing metrics, including average reward, collision rate, and average velocity, are used for a comprehensive evaluation.
\subsection{Training Process}
\begin{table*}[htbp]
    \centering
    \renewcommand{\arraystretch}{1.1}
    \setlength{\tabcolsep}{8.5pt}
    \caption{Policy Evaluation Results}
    \small
    \begin{tabular}{lccccccccc}
        \toprule
        \multirow{2}{*}{\textbf{Models}} & \multicolumn{3}{c}{\textbf{Turn Left}} & \multicolumn{3}{c}{\textbf{Go Straight}} & \multicolumn{3}{c}{\textbf{Turn Right}} \\
        \cmidrule(lr){2-4} \cmidrule(lr){5-7} \cmidrule(lr){8-10}
        & C.R. (\%) & A.R.  & A.V. (m/s) 
        & C.R. (\%) & A.R.  & A.V. (m/s)  
        & C.R. (\%) & A.R.  & A.V. (m/s) \\
        \midrule
        GCN-DQN  & 27.40  & 4.79 & 7.71    & 15.45  & 5.48 & 7.69   & 12.40  & 5.86 & 8.60   \\
        
        GCN-Double-DQN    & 24.65  & 4.96 & 7.72    & 18.40  & 5.27 & 7.75    & 12.05  & 6.05 & 8.45   \\
       
        GCN-Dueling-DQN  & 23.25  & 5.05 & 7.70    & 16.60  & 5.39 & 7.71    & 9.80  & 6.74 & 8.37   \\
       
        GCN-D3QN  &22.35 &5.13 &7.76     & 14.20  & 5.56 & 7.73          & 9.20 & 6.98 & 8.41  \\
        
        GAT-D3QN  & 19.40  & 5.25 & 7.74       & 17.00  & 5.35 & 7.78         & 11.35  & 6.21 & 8.28  \\
        
        GCN-GAT-D3QN  & 20.15  & 5.36 & 7.87      & 14.45  & 5.64 & 7.85        & 10.25  & 6.46 & 8.31   \\
        
        \textbf{CGRL(ours)} & \textbf{13.50} & \textbf{5.96} & \textbf{7.81}  
        & \textbf{11.35} & \textbf{6.12} & \textbf{7.84}  
        & \textbf{7.00} & \textbf{7.42} & \textbf{8.46} \\
        \bottomrule
    \end{tabular}
    \label{tab:performance}
\end{table*}
This subsection outlines the training procedures for all decision-making algorithms across three driving tasks: turning left, turning right, and going straight. In Figure 3, it shows that the average rewards obtained by each algorithm progressively increase and eventually converge to a stable range across all tasks. During training, the average reward and episode length increase progressively until a stable convergence state is reached. Among the compared algorithms, the CGRL algorithm demonstrates superior learning efficiency over other baseline algorithms.
\\The integration of causal learning with GRL is essential for effectively capturing causal features and enhancing overall system performance. This efficiency is further enhanced by CGRL’s ability to extract causal representations from the interaction states of AEVs, while disregarding non-causal features that may impair decision-making. Moreover, by isolating these causal features, CGRL facilitates the learning of invariant representations, thereby mitigating data inefficiency challenges commonly encountered in GRL frameworks.
\subsection{Performance Evaluation}
Table III reveals that the left-turning and straight-driving tasks exhibit higher collision rates across all decision-making algorithms when compared to the right-turning scenario. This can be attributed to the increased complexity of these maneuvers, which require the AEV to carefully observe its surroundings, assess traffic dynamics, and select appropriate actions to navigate through more intricate interactions. In contrast, right-turning is a relatively simpler maneuver, as it typically involves fewer conflicts with other vehicles, thereby reducing the likelihood of collisions. \\
The proposed CGRL algorithm outperforms all baseline methods across multiple performance metrics over 2000 testing episodes, demonstrating its effectiveness in complex intersection navigation scenarios.
One key indicator of this performance is the average velocity, which reflects the trade-off between driving efficiency and safety. CGRL achieves well-balanced average velocities of 7.81 m/s for left turns, 7.84 m/s for straight driving, and 8.46 m/s for right turns. While these values is not able to represent the highest speeds among all compared models, they indicate that CGRL maintains an optimal balance, allowing the AEV to make appropriate, safe decisions while ensuring smooth and timely intersection crossings. \\
Another critical metric is the average reward, which captures the algorithm’s overall effectiveness in achieving the driving task objectives, such as minimizing intersection traversal time and promoting smooth, efficient navigation. CGRL attains the highest average reward across all tested scenarios, left-turning, right-turning, and going straight. This superior performance demonstrates CGRL's ability to learn optimal control policies that balance speed and safety, highlighting its effectiveness in maintaining efficient traffic flow while ensuring that the AEV operates within safe behavioral margins.\\
CGRL also achieves significantly lower collision rates compared to the baseline models, underscoring its strong emphasis on safety. Specifically, it achieves the lowest collision rates for all maneuver types: 13.50 percent for left turns, 11.35 percent for straight driving, and 7.00 percent for right turns. These results suggest that CGRL enables the AEV to make safer decisions, particularly in high-interaction, multi-agent environments. Therefore, the effectiveness of CGRL in achieving both safety and efficiency can be attributed to its unique integration of causal learning and GRL. By explicitly modeling interactions among surrounding vehicles using graph structures, CGRL enables a fine-grained understanding of inter-agent relationships and dynamics. The decision-making framework leverages this structural information to extract causal features, thereby facilitating more informed and reliable decisions. As a result, CGRL prioritizes collision avoidance while minimizing disruptions to traffic flow, ultimately achieving superior performance across a range of challenging driving tasks.

\section{Conclusion}
In this study, we propose a multi-agent decision-making framework CGRL, which combines causal learning with GRL, successfully identifies and leverages causal features that influence optimal decision-making in autonomous vehicles. This framework systematically identifies and exploits causal features that influence its decision-making processes, thereby facilitating more informed and optimized behaviors within multi-agent environments. Empirical evaluations reveal that the method attains superior average rewards throughout training and markedly surpasses baseline approaches across critical performance metrics, including collision rates and average cumulative rewards during testing. These findings underscore the potential of CGRL algorithm to enhance the safety and efficiency of autonomous vehicles in complex multi-agent environments.\\ Future work will implement real-world validation trials to examine the framework's performance and extend the current methodology to incorporate complex environmental interactions, particularly focusing on heterogeneous traffic participants.

\bibliographystyle{IEEEtran}

\bibliography{ref}

\begin{thebibliography}{10}
\providecommand{\url}[1]{#1}
\csname url@samestyle\endcsname
\providecommand{\newblock}{\relax}
\providecommand{\bibinfo}[2]{#2}
\providecommand{\BIBentrySTDinterwordspacing}{\spaceskip=0pt\relax}
\providecommand{\BIBentryALTinterwordstretchfactor}{4}
\providecommand{\BIBentryALTinterwordspacing}{\spaceskip=\fontdimen2\font plus
\BIBentryALTinterwordstretchfactor\fontdimen3\font minus \fontdimen4\font\relax}
\providecommand{\BIBforeignlanguage}[2]{{%
\expandafter\ifx\csname l@#1\endcsname\relax
\typeout{** WARNING: IEEEtran.bst: No hyphenation pattern has been}%
\typeout{** loaded for the language `#1'. Using the pattern for}%
\typeout{** the default language instead.}%
\else
\language=\csname l@#1\endcsname
\fi
#2}}
\providecommand{\BIBdecl}{\relax}
\BIBdecl

\bibitem{Adibi2023GraphNN}
P.~Adibi, B.~Shoushtarian, and J.~Chanussot, ``Graph neural networks and reinforcement learning: A survey,'' 2023.

\bibitem{10161704}
S.~Munikoti, D.~Agarwal, L.~Das, M.~Halappanavar, and B.~Natarajan, ``Challenges and opportunities in deep reinforcement learning with graph neural networks: A comprehensive review of algorithms and applications,'' \emph{IEEE Transactions on Neural Networks and Learning Systems}, vol.~35, no.~11, pp. 15\,051--15\,071, 2024.

\bibitem{Velickovic2017GraphAN}
P.~Velickovic, G.~Cucurull, A.~Casanova, A.~Romero, P.~Lio’, and Y.~Bengio, ``Graph attention networks,'' \emph{ArXiv}, vol. abs/1710.10903, 2017.

\bibitem{Kipf2016SemiSupervisedCW}
T.~Kipf and M.~Welling, ``Semi-supervised classification with graph convolutional networks,'' \emph{ArXiv}, vol. abs/1609.02907, 2016.

\bibitem{Job2023ExploringCL}
S.~Job, X.~Tao, T.~Cai, H.~Xie, L.~Li, J.~Yong, and Q.~Li, ``Exploring causal learning through graph neural networks: An in-depth review,'' \emph{ArXiv}, vol. abs/2311.14994, 2023.

\bibitem{Jiang2023WhenGN}
W.~Jiang, H.~Liu, and H.~Xiong, ``When graph neural network meets causality: Opportunities, methodologies and an outlook,'' 2023.

\bibitem{Zeng2023ASO}
Y.~Zeng, R.~Cai, F.~Sun, L.~Huang, and Z.~Hao, ``A survey on causal reinforcement learning,'' \emph{ArXiv}, vol. abs/2302.05209, 2023.

\bibitem{Deng2023CausalRL}
Z.-H. Deng, J.~Jiang, G.~Long, and C.~Zhang, ``Causal reinforcement learning: A survey,'' \emph{Trans. Mach. Learn. Res.}, vol. 2023, 2023.

\bibitem{Grimbly2021CausalMR}
S.~J. Grimbly, J.~Shock, and A.~Pretorius, ``Causal multi-agent reinforcement learning: Review and open problems,'' \emph{ArXiv}, vol. abs/2111.06721, 2021.

\bibitem{Kipf2016VariationalGA}
T.~Kipf and M.~Welling, ``Variational graph auto-encoders,'' \emph{ArXiv}, vol. abs/1611.07308, 2016.

\bibitem{sai2024}
S.~Munikoti, D.~Agarwal, L.~Das, M.~Halappanavar, and B.~Natarajan, ``Challenges and opportunities in deep reinforcement learning with graph neural networks: A comprehensive review of algorithms and applications,'' \emph{IEEE Transactions on Neural Networks and Learning Systems}, vol.~35, no.~11, pp. 15\,051--15\,071, 2024.

\bibitem{Liu2022GraphCD}
Q.~Liu, Z.~Li, X.~Li, J.~Wu, and S.~Yuan, ``Graph convolution-based deep reinforcement learning for multi-agent decision-making in mixed traffic environments,'' \emph{ArXiv}, vol. abs/2201.12776, 2022.

\bibitem{Yang2022GeneralizedSG}
F.~Yang, X.~Li, Q.~Liu, Z.~Li, and X.~Gao, ``Generalized single-vehicle-based graph reinforcement learning for decision-making in autonomous driving,'' \emph{Sensors (Basel, Switzerland)}, vol.~22, 2022.

\bibitem{Gao2022MultiAgentDM}
X.~Gao, X.~Li, Q.~Liu, Z.-H. Li, F.~Yang, and T.~Luan, ``Multi-agent decision-making modes in uncertain interactive traffic scenarios via graph convolution-based deep reinforcement learning,'' \emph{Sensors (Basel, Switzerland)}, vol.~22, 2022.

\bibitem{Chen2021GraphNN}
S.~Chen, J.~Dong, P.~Y.~J. Ha, Y.~Li, and S.~Labi, ``Graph neural network and reinforcement learning for multi‐agent cooperative control of connected autonomous vehicles,'' \emph{Computer‐Aided Civil and Infrastructure Engineering}, vol.~36, pp. 838 -- 857, 2021.

\bibitem{Klimke2022CooperativeBP}
M.~Klimke, B.~V{\"o}lz, and M.~Buchholz, ``Cooperative behavioral planning for automated driving using graph neural networks,'' \emph{ArXiv}, vol. abs/2202.11376, 2022.

\bibitem{Shi2020EfficientCA}
T.~Shi, J.~Wang, Y.~Wu, L.~Miranda-Moreno, and L.~Sun, ``Efficient connected and automated driving system with multi-agent graph reinforcement learning,'' 2020.

\bibitem{Cai2021DQGATTS}
P.~Cai, H.~Wang, Y.~Sun, and M.~Liu, ``Dq-gat: Towards safe and efficient autonomous driving with deep q-learning and graph attention networks,'' \emph{IEEE Transactions on Intelligent Transportation Systems}, vol.~23, pp. 21\,102--21\,112, 2021.

\bibitem{DRL-GAT-SA}
\BIBentryALTinterwordspacing
Y.~Peng, G.~Tan, H.~Si, and J.~Li, ``Drl-gat-sa: Deep reinforcement learning for autonomous driving planning based on graph attention networks and simplex architecture,'' \emph{J. Syst. Archit.}, vol. 126, no.~C, May 2022. [Online]. Available: \url{https://doi.org/10.1016/j.sysarc.2022.102505}
\BIBentrySTDinterwordspacing

\bibitem{MndezMolina2020CausalBQ}
A.~M{\'e}ndez-Molina, I.~Feliciano-Avelino, E.~F. Morales, and L.~E. Sucar, ``Causal based q-learning,'' \emph{Res. Comput. Sci.}, vol. 149, pp. 95--104, 2020.

\bibitem{Lu2022EfficientRL}
Y.~Lu and A.~Tewari, ``Efficient reinforcement learning with prior causal knowledge,'' in \emph{CLEaR}, 2022.

\bibitem{Hart2020CounterfactualPE}
P.~Hart and A.~Knoll, ``Counterfactual policy evaluation for decision-making in autonomous driving,'' \emph{arXiv: Learning}, 2020.

\bibitem{Peng2022CausalitydrivenHS}
S.~Peng, X.~Hu, R.~Zhang, K.~Tang, J.~Guo, Q.~Yi, R.~Chen, X.~Zhang, Z.~Du, L.~Li, Q.~Guo, and Y.~Chen, ``Causality-driven hierarchical structure discovery for reinforcement learning,'' \emph{ArXiv}, vol. abs/2210.06964, 2022.

\bibitem{Spirtes2000ConstructingBN}
P.~Spirtes, C.~Glymour, R.~Scheines, S.~A. Kauffman, V.~Aimale, and F.~C. Wimberly, ``Constructing bayesian network models of gene expression networks from microarray data,'' 2000.

\bibitem{alali:hal-04666466}
\BIBentryALTinterwordspacing
S.~Al-Ali and I.~Balelli, ``{Multi-Channel Causal Variational Autoencoder},'' Aug. 2024, working paper or preprint. [Online]. Available: \url{https://hal.science/hal-04666466}
\BIBentrySTDinterwordspacing

\bibitem{Reddy2021OnCD}
A.~G. Reddy, B.~G. L, and V.~N. Balasubramanian, ``On causally disentangled representations,'' in \emph{AAAI Conference on Artificial Intelligence}, 2021.

\bibitem{Shen2020WeaklySD}
X.~Shen, F.~Liu, H.~Dong, Q.~Lian, Z.~Chen, and T.~Zhang, ``Weakly supervised disentangled generative causal representation learning,'' \emph{J. Mach. Learn. Res.}, vol.~23, pp. 241:1--241:55, 2020.

\bibitem{Yang2020CausalVAEDR}
M.~Yang, F.~Liu, Z.~Chen, X.~Shen, J.~Hao, and J.~Wang, ``Causalvae: Disentangled representation learning via neural structural causal models,'' \emph{2021 IEEE/CVF Conference on Computer Vision and Pattern Recognition (CVPR)}, pp. 9588--9597, 2020.

\bibitem{Feng2023ConceptfreeCD}
J.~Feng, L.~Zhang, and L.~Yang, ``Concept-free causal disentanglement with variational graph auto-encoder,'' \emph{ArXiv}, vol. abs/2311.10638, 2023.

\bibitem{Pourkeshavarz2024CaDeTAC}
M.~Pourkeshavarz, J.~Zhang, and A.~Rasouli, ``Cadet: A causal disentanglement approach for robust trajectory prediction in autonomous driving,'' \emph{2024 IEEE/CVF Conference on Computer Vision and Pattern Recognition (CVPR)}, pp. 14\,874--14\,884, 2024.

\bibitem{Brody2021HowAA}
S.~Brody, U.~Alon, and E.~Yahav, ``How attentive are graph attention networks?'' \emph{ArXiv}, vol. abs/2105.14491, 2021.

\bibitem{pmlr-v48-wangf16}
Z.~Wang, T.~Schaul, M.~Hessel, H.~Hasselt, M.~Lanctot, and N.~Freitas, ``Dueling network architectures for deep reinforcement learning,'' in \emph{Proceedings of The 33rd International Conference on Machine Learning}, ser. Proceedings of Machine Learning Research, M.~F. Balcan and K.~Q. Weinberger, Eds., vol.~48.\hskip 1em plus 0.5em minus 0.4em\relax New York, New York, USA: PMLR, 20--22 Jun 2016, pp. 1995--2003.

\bibitem{Ay2008InformationFI}
N.~Ay and D.~Polani, ``Information flows in causal networks,'' \emph{Adv. Complex Syst.}, vol.~11, pp. 17--41, 2008.

\bibitem{Zheng2023CIGNNAG}
K.~Zheng, S.~Yu, and B.~Chen, ``Ci-gnn: A granger causality-inspired graph neural network for interpretable brain network-based psychiatric diagnosis,'' \emph{Neural networks : the official journal of the International Neural Network Society}, vol. 172, p. 106147, 2023.

\bibitem{Lin2022OrphicXAC}
W.~Lin, H.~Lan, H.~Wang, and B.~Li, ``Orphicx: A causality-inspired latent variable model for interpreting graph neural networks,'' \emph{2022 IEEE/CVF Conference on Computer Vision and Pattern Recognition (CVPR)}, pp. 13\,719--13\,728, 2022.

\bibitem{Pl2010EstimationOR}
D.~P{\'a}l, B.~P{\'o}czos, and C.~Szepesvari, ``Estimation of renyi entropy and mutual information based on generalized nearest-neighbor graphs,'' in \emph{Neural Information Processing Systems}, 2010.

\bibitem{Belghazi2018MINEMI}
I.~Belghazi, S.~Rajeswar, A.~Baratin, R.~D. Hjelm, and A.~C. Courville, ``Mine: Mutual information neural estimation,'' \emph{ArXiv}, vol. abs/1801.04062, 2018.

\bibitem{SanchezGiraldo2012MeasuresOE}
L.~G.~S. Giraldo, M.~Rao, and J.~C. Pr{\'i}ncipe, ``Measures of entropy from data using infinitely divisible kernels,'' \emph{IEEE Transactions on Information Theory}, vol.~61, pp. 535--548, 2012.

\bibitem{highway-env}
E.~Leurent, ``An environment for autonomous driving decision-making,'' \url{https://github.com/eleurent/highway-env}, 2018.

\bibitem{Treiber2000CongestedTS}
M.~Treiber, A.~Hennecke, and D.~Helbing, ``Congested traffic states in empirical observations and microscopic simulations,'' \emph{Physical review. E, Statistical physics, plasmas, fluids, and related interdisciplinary topics}, vol. 62 2 Pt A, pp. 1805--24, 2000.

\bibitem{Zhou2024ReasoningGR}
D.~Zhou, P.~Hang, and J.~Sun, ``Reasoning graph-based reinforcement learning to cooperate mixed connected and autonomous traffic at unsignalized intersections,'' \emph{Transportation Research Part C: Emerging Technologies}, 2024.

\end{thebibliography}

\end{document}